\documentclass[12pt]{article}

\usepackage{amsmath, amssymb, slashed}
\usepackage[dvips]{graphicx}
\usepackage{epsf}

\begin{document}

\begin{titlepage}
\begin{center}

\hfill ICRR-report-598-2011-15 \\
\hfill IPMU11-0177 \\

\vspace{2.0cm}
{\large\bf The GeV-scale dark matter with B$-$L asymmetry}
\vspace{2.0cm}

{\bf Masahiro Ibe}$^{(a,b)}$,
{\bf Shigeki Matsumoto}$^{(b)}$,
and
{\bf Tsutomu T. Yanagida}$^{(b)}$

\vspace{1.5cm}
{\it
$^{(a)}${\it ICRR, University of Tokyo, Kashiwa, 277-8582, Japan} \\
$^{(b)}${\it IPMU, TODIAS, University of Tokyo, Kashiwa, 277-8583, Japan} \\
}
\vspace{3.0cm}

\abstract{
One of important properties of dark matter is its stability. The U(1)$_{\rm B-L}$ gauge symmetry is the most attractive symmetry to guarantee the stability. Though the symmetry is expected to be broken at very high energy scale to account for tiny neutrino masses through the seesaw mechanism, the residual discrete symmetry of U(1)$_{\rm B-L}$ can stabilize the dark matter naturally. We prove that, when there is new physics connecting B$-$L charges of dark matter and standard model particles at the scale between the electroweak and the U(1)$_{\rm B-L}$ breaking, the mass of dark matter is definitely predicted to be (5--7)/$Q_{\rm DM}$ GeV ($Q_{DM}$ is the B$-$L charge of dark matter) independent not only of details of the new physics but also of its energy scale. We also show two attractive examples. First one is the scalar dark matter with the B$-$L charge of one, which turns out to be very consistent with current CoGeNT results. Another one is the fermionic dark matter having the B$-$L charge of one third, which is also attractive from the viewpoint of model building.}

\end{center}
\end{titlepage}
\setcounter{footnote}{0}

\section{Introduction}
\label{sec: intro}

There are many compelling evidences for the existence of dark matter in our universe. Some properties of the dark matter have also been clarified thanks to recent cosmological observations; the dark matter should be a neutral and stable (or long-lived) particle whose interactions are weaker or shorter-ranged than those of standard model (SM) particles. In addition, its cosmological abundance is observed to be $\Omega_{\rm DM} h^2 \simeq 0.110$~\cite{Komatsu:2010fb} and its motion at the epoch of large-scale structure formation of the universe turns out to be non-relativistic~\cite{Evrard:2001hu}. Among those properties, the most remarkable one is the stability from the viewpoint of particle physics.

The most promising way to stabilize dark matter is the use of the U(1)$_{\rm B-L}$ gauge symmetry~\cite{Nakayama:2011dj}. This symmetry requires right-handed neutrinos because of anomaly cancellation, and is expected to be broken at high energy scale of ${\cal O}(10^{12})$ GeV. Right-handed neutrinos therefore acquires their masses of this scale, which allow us to explain tiny neutrino masses through the seesaw mechanism~\cite{seesaw} and to generate the baryon asymmetry of the universe through the leptogenesis~\cite{leptogenesis}. Since Majorana mass terms of right-handed neutrinos have the B$-$L charge of two, a residual discrete symmetry of U(1)$_{\rm B-L}$ exists, which can be used to stabilize dark matter.

In the early universe, the decay of right-handed neutrinos is expected to produce the B$-$L asymmetry in the SM sector. Since the dark matter stabilized by U(1)$_{\rm B-L}$ carries the B$-$L charge, there is no wonder if the B$-$L asymmetry of the dark matter is also generated by the leptogenesis or some other mechanisms~\cite{Darkogenesis}. When the annihilation cross section between dark and anti-dark matters is large enough, the symmetric component of dark matter relics is eliminated and the B$-$L asymmetry of the dark matter gives the relic abundance of dark matter observed today. If new physics at the scale between right-handed neutrinos and SM particles does not exist, the amount of the B$-$L asymmetry in the dark matter sector as well as that in the SM sector depends on details of the decay of right-handed neutrinos. It is then difficult to give a definite prediction on these asymmetries. On the other hand, when there is the new physics connecting B$-$L charges of dark matter and SM particles, it is possible to give a definite prediction without depending not only on details of the new physics but also on its energy scale. We prove this remarkable result in the next section (Sec.\ref{sec: general prediction}) and show that the mass of dark matter is indeed definitely predicted to be (5--7)/$Q_{\rm DM}$ GeV with $Q_{DM}$ being the B$-$L charge of the dark matter.

According to the result obtained in Sec.\ref{sec: general prediction}, we also show two attractive examples of the dark matter in Sec.\ref{sec: examples}. First one is the scalar dark matter with the B$-$L charge of one. It turns out that the mass of the dark matter is predicted to be 5--7 GeV. Furthermore, when we require the annihilation cross section between dark and anti-dark matter particles to be large enough in order to eliminate their symmetric components, the scattering cross section between the dark matter and a nucleon is predicted to be ${\cal O}(10^{-4})$ pb, which is very consistent with the CoGeNT anomaly~\cite{CoGeNT}. Another example is the fermionic dark matter having the B$-$L charge of one third, whose mass is predicted to be 15--21 GeV. It turns out that a UV completed theory for the dark matter can be easily constructed with being consistent with several severe constraints that usually we have to worry about on dark matter candidates of this kind. Section \ref{sec: conclusion} is devoted to summary of our discussions.

\section{B$-$L charged dark matter}
\label{sec: general prediction}

Here, we show that the residual discrete symmetry of gauged U(1)$_{\rm B-L}$ is possible to guarantee the stability of dark matter naturally~\cite{Nakayama:2011dj}. The dark matter of this kind may have asymmetry in the universe. When new physics exists at the scale between electroweak and U(1)$_{\rm B-L}$ breaking, the dark matter could be in chemical equilibrium with SM particles through interactions connecting the asymmetries of the dark matter and SM particles. We show, if it is the case, that the ratio of asymmetries between baryon and dark matter observed today can be predicted without depending not only on details of the interactions but also on the scale of the new physics.

\subsection{Stability of the dark matter}

\begin{table}[t]
\centering
\begin{tabular}{c|cccccc|c}
& $Q$ & $u_R$ & $d_R$ & $L$ & $e_R$ & $H$ & Dark matter \\
\hline
SU(3)$_c$ & {\bf 3} & {\bf 3} & {\bf 3} & {\bf 1} & {\bf 1} & {\bf 1} & {\bf 1} \\
SU(2)$_L$ & {\bf 2} & {\bf 1} & {\bf 1} & {\bf 2} & {\bf 1} & {\bf 1} & {\bf 1} \\
U(1)$_Y$ & $1/6$ & $2/3$ & $-1/3$ & $-1/2$ & $-1$ & $1/2$ & $0$ \\
U(1)$_{\rm B-L}$ & $1/3$ & $1/3$ & $1/3$ & $-1$ & $-1$ & $0$ & $Q_{\rm DM}$ \\
\hline
\end{tabular}
\caption{\small Gauge quantum numbers of SM particles and dark matter.}
\label{tab: quantum numbers}
\end{table}

The existence of the U(1)$_{\rm B-L}$ gauge symmetry is very likely present from the viewpoints of neutrino masses~\cite{seesaw}, leptogenesis~\cite{leptogenesis}, and grand unification. Gauge quantum numbers of SM particles and dark matter are shown in Table~\ref{tab: quantum numbers}, where $Q = (u_L, b_L)^T$ and $u_R$ ($d_R$) are left- and right-handed quarks, while  $L = (\nu_L, e_L)^T$ and $e_R$ are left- and charged right-handed leptons, respectively. Higgs boson is denoted by $H$. Dark matter is assumed to be singlet under SM gauge groups to satisfy severe constraints form direct detection experiments of dark matter. We postulate that the dark matter is either a complex scalar or a Dirac fermion in this article.

Though quarks have the B$-$L charge of one third, they are confined into baryons and mesons with the charges of one and zero, respectively. The SM is therefore composed only of fermions with odd B$-$L charge and bosons with even B$-$L charge. On the other hand, since U(1)$_{\rm B-L}$ is spontaneously broken with an vacuum expectation value with the charge of two, the residual $Z_2$ symmetry exists and it can be used to stabilize dark matter. For instance, scalar with odd B$-$L charge or fermion with even B$-$L charge is stable~\cite{Nakayama:2011dj}. There is another possibility. When dark matter has a fractional B$-$L charge, for example, $Q_{\rm DM} =$ 1/2, 1/3, $\cdots$ 1/n, the residual symmetry is enhanced to $Z_{2n}$, which is also possible to stabilize dark matter.

\subsection{Baryon and dark matter asymmetries in the universe}

We next consider how the dark matter charged under U(1)$_{\rm B-L}$ behaves in the early universe. In general, the decay of right-handed neutrinos produces the B$-$L asymmetries both in the dark matter sector and the SM sector. If there is no new physics connecting B$-$L charges of dark matter to those of SM particles at the scale between electroweak and right-handed neutrinos (the breaking scale of U(1)$_{\rm B-L}$), the produced asymmetries are determined by complicated non-equilibrium processes~\cite{Darkogenesis}, which makes us difficult to have a definite prediction on these asymmetries. On the other hand, when such a new physics exists, the dark matter is expected to be in chemical equilibrium with SM particles through following interactions,
\begin{eqnarray}
{\cal L}_{\rm int}
\simeq
\frac{1}{\Lambda^n}{\cal O}_{\rm DM} \cdot {\cal O}_{\rm SM} + c.c.,
\label{eq: ADM interaction}
\end{eqnarray}
where ${\cal O}_{\rm DM}$ involves only dark(anti-dark) matter field, while ${\cal O}_{\rm SM}$ consists of SM fields. The scale of new physics is denoted by $\Lambda$ and the interaction is assumed to be the lowest dimensional one breaking the dark matter number, but preserving the total B$-$L number. The baryon asymmetry of the universe is, as a result, related to the B$-$L asymmetry of the dark matter through the interaction as long as $\Lambda$ satisfies,
\begin{eqnarray}
\Lambda
\lesssim
T_{\rm lept} \left( \frac{M_{\rm pl}}{T_{\rm lept}} \right)^{1/(2n)},
\end{eqnarray}
where $T_{\rm lept}$ is the temperature that the leptogenesis occurs, which is given by the decay temperature of right-handed neutrinos. Planck scale is denoted by $M_{\rm pl} \simeq 2.43 \times 10^{18}$ GeV. This is exactly the same as the mechanism used in the asymmetric dark matter scenario~\cite{Kaplan:2009ag}. Interestingly, it is possible to predict the ratio of these asymmetries without knowing details of the interaction as we will see below.

\subsubsection{Case I ($T_D > T_{sph}$)}

Here, we consider the case that the interaction in eq.(\ref{eq: ADM interaction}), where we call it the ADM interaction, decouples before the Sphaleron process decouples. We use the character $T_D$ to represent the decoupling temperature of the ADM interaction, while $T_{sph}$ is for that of Sphaleron which is estimated to be $T_{sph} \simeq [80 + 54 (m_h/{\rm 120 GeV})]$ GeV for a given higgs mass $m_h$~\cite{AristizabalSierra:2010mv}. When the temperature of the universe is below $T_D$, both dark matter asymmetry and B$-$L asymmetry of the SM sector are individually conserved. When the temperature becomes lower than $T_{sph}$, all of dark matter asymmetry, B and L asymmetries of the SM sector are individually conserved.

Let us consider the relation of chemical potentials when the temperature of the universe is around $T_D$. Since the chemical potentials of gauge bosons are zero before the electroweak symmetry breaking, we first consider the potentials of $Q$, $u_R$, $d_R$, $L$, $e_R$, and $H$, which are denoted by $\mu_Q$, $\mu_{u_R}$, $\mu_{d_R}$, $\mu_L$, $\mu_{e_R}$, and $\mu_H$, respectively. From Yukawa interactions and Sphaleron process, we obtain following relations~\cite{Harvey:1990qw},
\begin{eqnarray}
-\mu_Q - \mu_H + \mu_{u_R} &=& 0, \label{eq: QHU} \\
-\mu_Q + \mu_H + \mu_{d_R} &=& 0, \label{eq: QHD} \\
-\mu_L + \mu_H + \mu_{e_R} &=& 0, \label{eq: LHE} \\
3\mu_Q + \mu_L &=& 0. \label{eq: Sphaleron 1}
\end{eqnarray}
Other SM interactions such as gauge interactions and self-interactions of higgs bosons do not give additional information for the chemical potentials.

We next consider the relation of chemical potentials obtained from the ADM interaction in eq.(\ref{eq: ADM interaction}). Suppose that the operator ${\cal O}_{\rm DM}$ has the dark matter number of $N_{\rm DM}$, meaning the number of dark matter field minus that of anti-dark matter field is $N_{\rm DM}$, so that $N_{\rm DM}$ is given by the B$-$L charge of the operator ${\cal O}_{\rm DM}$ through the equation, $Q_{\rm B-L}({\cal O}_{\rm DM}) = Q_{\rm DM} \times N_{\rm DM}$ with $Q_{\rm DM}$ being the B$-$L charge of the dark matter. On the other hand, ${\cal O}_{\rm SM}$ has $N_Q$, $N_{u_R}$, $N_{d_R}$, $N_L$, $N_{e_R}$, and $N_H$ numbers for $Q$, $u_R$, $d_R$, $L$, $e_R$, and $H$ fields, respectively. We then have the relation,
\begin{eqnarray}
N_{\rm DM}\mu_{\rm DM} + N_Q \mu_Q + N_{u_R} \mu_{u_R} + N_{d_R} \mu_{d_R}
+ N_L \mu_L + N_{e_R} \mu_{e_R} + N_H \mu_H = 0,
\label{eq: ADM 1}
\end{eqnarray}
where $\mu_{\rm DM}$ is the chemical potential of the dark matter. The ADM interaction should be singlet under SM gauge interactions. In addition, it must be also singlet under U(1)$_{\rm B-L}$, otherwise B$-$L asymmetry produced in the very early universe is washed-out. As a result, we have following constraints; $N_Q/6 + 2N_{u_R}/3 - N_{d_R}/3 - N_L/2 - N_{e_R} + N_H/2 = 0$ and $N_Q/3 + N_{u_R}/3 + N_{d_R}/3 - N_L - N_{e_R} + Q_{\rm DM} N_{\rm DM} = 0$. With the use of these relations and those in eq.(\ref{eq: QHU})--(\ref{eq: ADM 1}), $\mu_{\rm DM}$ turns out to be
\begin{eqnarray}
\mu_{\rm DM} = -\frac{11}{7}Q_{\rm DM} \mu_L.
\label{eq: DM 1}
\end{eqnarray}
It is very interesting to see that $\mu_{\rm DM}$ does not depend on details of the ADM interaction ($N_{\rm DM}$, $N_Q$, and so on) and does depend only on the B$-$L charge $Q_{\rm DM}$.

In addition to the above relations between chemical potentials, we also have another relation obtained by the neutrality of the universe, which is given by
\begin{eqnarray}
6\mu_{u_R} + 3\mu_Q - 3\mu_{d_R} - 3\mu_L - 3\mu_{e_R} + 2n_H\mu_H = 0.
\label{eq: neutrality}
\end{eqnarray}
Since we do not know the higgs sector of the SM, we simply postulate that it is composed of $n_H$ higgs boson doublets. It is also possible to add singlet higgs bosons to the sector, which does not alter above discussions because singlet higgs bosons are electrically neutral and do not have non-zero chemical potentials.

Using eqs. (\ref{eq: QHU}), (\ref{eq: QHD}), (\ref{eq: LHE}), (\ref{eq: Sphaleron 1}), (\ref{eq: DM 1}), and (\ref{eq: neutrality}), the ratio of B$-$L asymmetry of the SM sector, $({\rm B} - {\rm L})_{\rm SM}$, and that of dark matter, $({\rm B} - {\rm L})_{\rm DM}$, is predicted to be
\begin{eqnarray}
\frac{({\rm B} - {\rm L})_{\rm SM}}{\rm ({\rm B} - {\rm L})_{\rm DM}}
=
\frac{7(66 + 13n_H)}{22(6 + n_H)Q_{\rm DM}^2},
\label{eq: B-L vs DM}
\end{eqnarray}
when the temperature of the universe is around $T_D$. The asymmetry $({\rm B} - {\rm L})_{\rm SM}$ is divided into baryon and lepton asymmetries of the SM sector (B$_{\rm SM}$ and L$_{\rm SM}$) when the Sphaleron process decouples, where B$_{\rm SM}$ is nothing but the baryon asymmetry observed today. The ratio between $({\rm B} - {\rm L})_{\rm SM}$ and B$_{\rm SM}$ is well known to be ${\rm B}_{\rm SM}/({\rm B} - {\rm L})_{\rm SM} = 30/97 \simeq 0.31$, and we finally obtain the ratio ${\rm B}_{\rm SM}/({\rm B} - {\rm L})_{\rm DM}$ as
\begin{eqnarray}
\frac{{\rm B}_{\rm SM}}{({\rm B} - {\rm L})_{\rm DM}}
=
\frac{105}{1067} \frac{66 + 13n_H}{6 + n_H}
\frac{1}{Q_{\rm DM}^2}.
\label{eq: general prediction 1}
\end{eqnarray}
In the calculation of ${\rm B}_{\rm SM}/({\rm B} - {\rm L})_{\rm SM}$, we have assumed that charged higgs bosons predicted by $n_H \geq 2$ higgs doublets have already been decoupled at $T_{sph}$.

When the annihilation cross section between dark and anti-dark matter particles is large enough and, as a result, the symmetric component of dark matter relics is eliminated in the universe, the B$-$L asymmetry of the dark matter is directly related to its observed abundance, as in the case of the baryon asymmetry. Dark matter and baryon asymmetries are then given by $({\rm B} - {\rm L})_{\rm DM} = \Omega_{\rm DM} Q_{\rm DM} \rho_c/(s_0 m_{\rm DM})$ and B$_{\rm SM}$ $= \Omega_b \rho_c/(s_0 m_N)$, where $\rho_c \simeq 1.05 \times 10^{-5} h^{-2}$ GeV/cm$^3$ and $s_0 \simeq 2890$ are critical energy and entropy densities of the present universe, while $\Omega_{\rm DM} h^2 \simeq 0.110$ and $\Omega_b h^2 \simeq 0.0227$ are density parameters of dark matter and baryon~\cite{Nakamura:2010zzi}. Dark matter and nucleon masses are denoted by $m_{\rm DM}$ and $m_N \simeq 938$ MeV. Using the prediction in eq.(\ref{eq: general prediction 1}), the mass of dark matter is now predicted in the following formula,
\begin{eqnarray}
m_{\rm DM}
=
\frac{105}{1067} \frac{66 + 13n_H}{6 + n_H}
\frac{\Omega_{\rm DM}}{\Omega_b} \frac{m_N}{Q_{\rm DM}},
\label{eq: DM mass 1}
\end{eqnarray}
which takes a value of $m_{\rm DM} \simeq 5.1/Q_{\rm DM}$ GeV for the cases of $n_H =$ 1 or 2.

\subsubsection{Case II ($T_D < T_{sph}$)}

Here, we consider the case that the Sphaleron process decouples before the ADM interaction decouples. In this case, when the temperature of the universe is below $T_{sph}$, both B and L asymmetries are preserved and the dark matter is regarded as a particle which carries B and/or L charge(s), depending on ${\cal O}_{\rm SM}$. As in the case of previous subsection, let us consider the relation of chemical potentials when the temperature is above $T_{sph}$. SM interactions and Sphaleron process lead to
\begin{eqnarray}
-\mu_\nu + \mu_W + \mu_e &=& 0, \label{eq: LCC} \\
-\mu_u + \mu_W + \mu_d &=& 0, \label{eq: QCC} \\
2\mu_u + \mu_d + \mu_e &=& 0, \label{eq: Sphaleron 2}
\end{eqnarray}
where $\mu_\nu$, $\mu_e$, $\mu_u$, $\mu_d$, and $\mu_W$ are chemical potentials of $\nu_L$, $e$ (composed of $e_L$ \& $e_R$), $u$ ($u_L$ \& $u_R$), $d$ ($d_L$ \& $d_R$), and $W$ boson, respectively. On the other hand, the relation obtained by the ADM interaction has, in general, the following from,
\begin{eqnarray}
N_{\rm DM}\mu_{\rm DM} + N_e \mu_e + N_\nu \mu_\nu
+ N_u \mu_u + N_d \mu_d + N_W \mu_W = 0,
\label{eq: ADM 2}
\end{eqnarray}
The requirement that the ADM interaction should be neutral under electromagnetic interaction and also singlet under U(1)$_{\rm B-L}$ leads to following constraints; $-N_e + 2N_u/3 - N_d/3 + N_W = 0$ and $-N_e - N_\nu + N_u/3 + N_d/3 + Q_{\rm DM}N_{\rm DM} = 0$. With the use of these relations and those in eq.(\ref{eq: LCC})--(\ref{eq: ADM 2}), $\mu_{\rm DM}$ turns out to be
\begin{eqnarray}
\mu_{\rm DM} = -Q_{\rm DM} \mu_L.
\label{eq: DM 2}
\end{eqnarray}
Again, the chemical potential $\mu_{\rm DM}$ does not depend on details of the ADM interaction ($N_{\rm DM}$, $N_e$, and so on). This time, the neutrality condition of the universe gives
\begin{eqnarray}
8\mu_u - 6\mu_d - 6\mu_e + 6\mu_W = 0,
\label{eq: neutrality 2}
\end{eqnarray}
where the contribution from the top quark to above equation has been negrected, because the mass of top quark is larger than the decoupling temperature $T_{sph}$.

After the Sphaleron process decouples, B and L asymmetries are preserved. Contributions to these asymmetries from dark matter sector are always constant unless the non-relativistic effect of dark matter mass is significant. As a result, the ratio between baryon asymmetry of the SM sector and that of dark matter is given by
\begin{eqnarray}
\frac{{\rm B}_{\rm SM}}{({\rm B} - {\rm L})_{\rm DM}}
=
\frac{4\mu_u + 6\mu_d}{2\mu_{\rm DM}}
=
\frac{45}{29}\frac{1}{Q_{\rm DM}^2}.
\label{eq: general prediction 2}
\end{eqnarray}
When the decoupling temperature is very low so that the dark matter is non-relativistic at $T_D$, above prediction is not valid anymore~\cite{Buckley:2010ui}. In such a case, it is difficult to obtain the prediction on ${\rm B}_{\rm SM}/{\rm DM}$ without depending on the details of the ADM interaction. When the decoupling temperature $T_D$ is high enough, the mass of dark matter is, with the use of the result in eq.(\ref{eq: general prediction 2}), predicted to be
\begin{eqnarray}
m_{\rm DM}
=
\frac{45}{29} \frac{\Omega_{\rm DM}}{\Omega_b} \frac{m_N}{Q_{\rm DM}},
\label{eq: DM mass 2}
\end{eqnarray}
which takes a value of $m_{\rm DM} \simeq 7.1/Q_{\rm DM}$ GeV. The difference between the predictions on the mass of dark matter in eqs.(\ref{eq: DM mass 1}) and (\ref{eq: DM mass 2}) comes from what kinds of elementally excitations, namely, particles composing thermal bath, we have at the decoupling temperature $T_D$. In former case, excitations are from SM particles before electroweak symmetry breaking, while from those after the breaking in latter case.

\section{Examples of the B$-$L charged dark matter}
\label{sec: examples}

In this section, we consider two attractive examples of models for the dark matter charged under U(1)$_{\rm B-L}$. One is for the bosonic dark matter with the B$-$L charge of one and another is for the fermionic dark matter with the charge of one third.

\subsection{Bosonic dark matter model}

Let us begin with the model for the bosonic dark matter. Although there are numerous operators ${\cal O}_{\rm SM}$ which are singlet under SM gauge group but carry non-vanishing B$-$L charges, the operators should be bosonic in this case. The lowest dimensional operator which is bosonic and charged under U(1)$_{\rm B-L}$ in the SM is ${\cal O}_{SM} = (LH)^2$. Therefore, for the bosonic dark matter with the B$-$L charge of one, the lowest dimensional connection between the dark matter and SM particles is given by,
\begin{eqnarray}
{\cal L}_{\rm int} = \frac{1}{\Lambda^3} \phi^2 (LH)^2.
\label{eq: bosonic connection}
\label{eq: bosonic}
\end{eqnarray}
According to the generic argument in previous section, the dark matter mass is predicted to be $m \simeq 5.1$ or $7.1$ GeV, depending on whether the decoupling temperature $T_D$ is above or below $T_{sph}$. The scale of new physics should satisfy the condition,
\begin{eqnarray}
m \left( \frac{M_{\rm pl}}{m} \right)^{1/6}
\ll
\, \Lambda \,
\lesssim
T_{\rm lept} \left( \frac{M_{\rm pl}}{T_{\rm lept}} \right)^{1/6},
\end{eqnarray}
where the lower bound on $\Lambda$ comes from the requirement that the asymmetry-transfer caused by the interaction in eq.(\ref{eq: bosonic connection}) ceases when the dark matter is relativistic.

\begin{figure}[t]
\begin{center}
\includegraphics[scale=0.35]{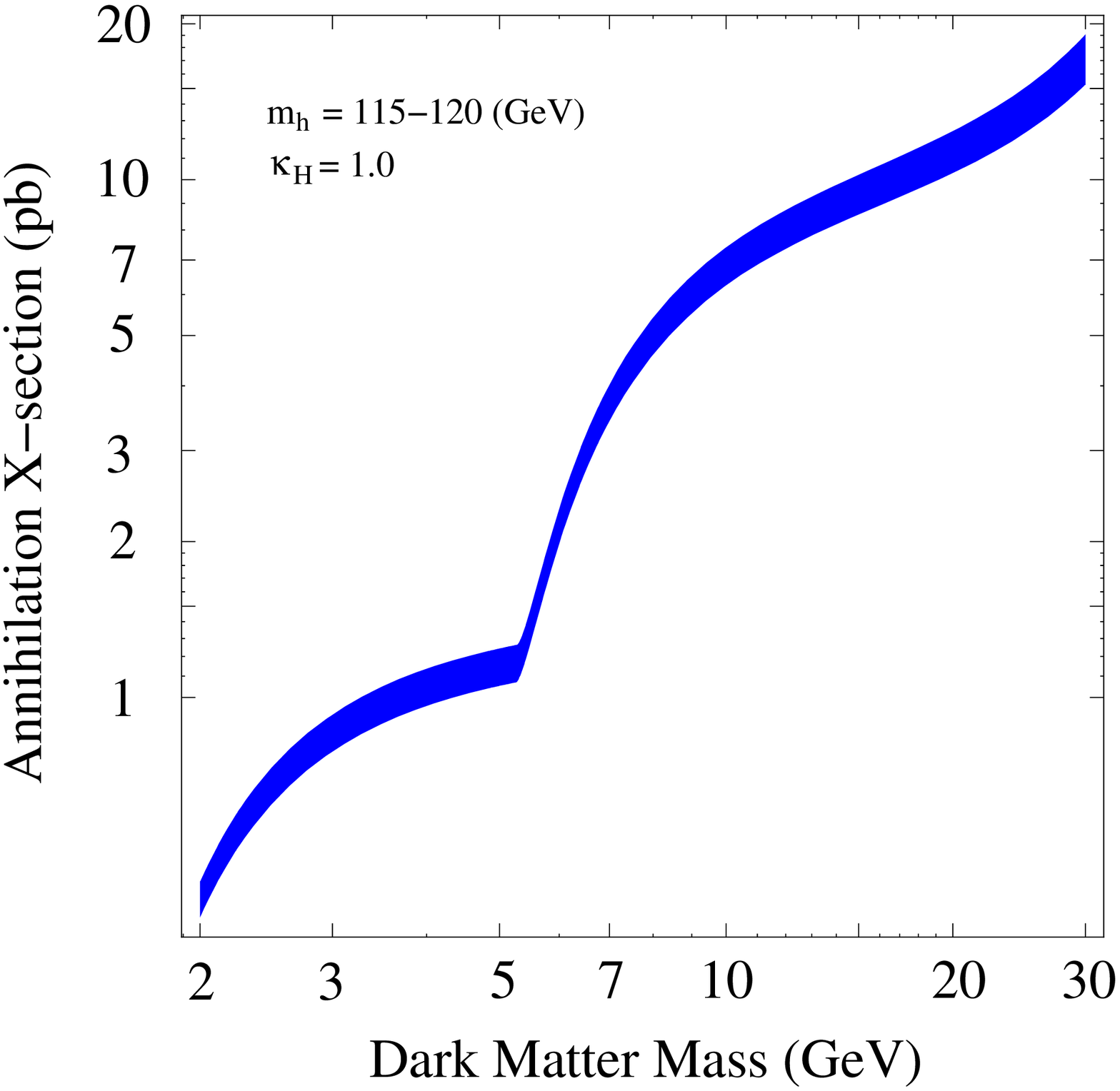}
~~~~~
\includegraphics[scale=0.35]{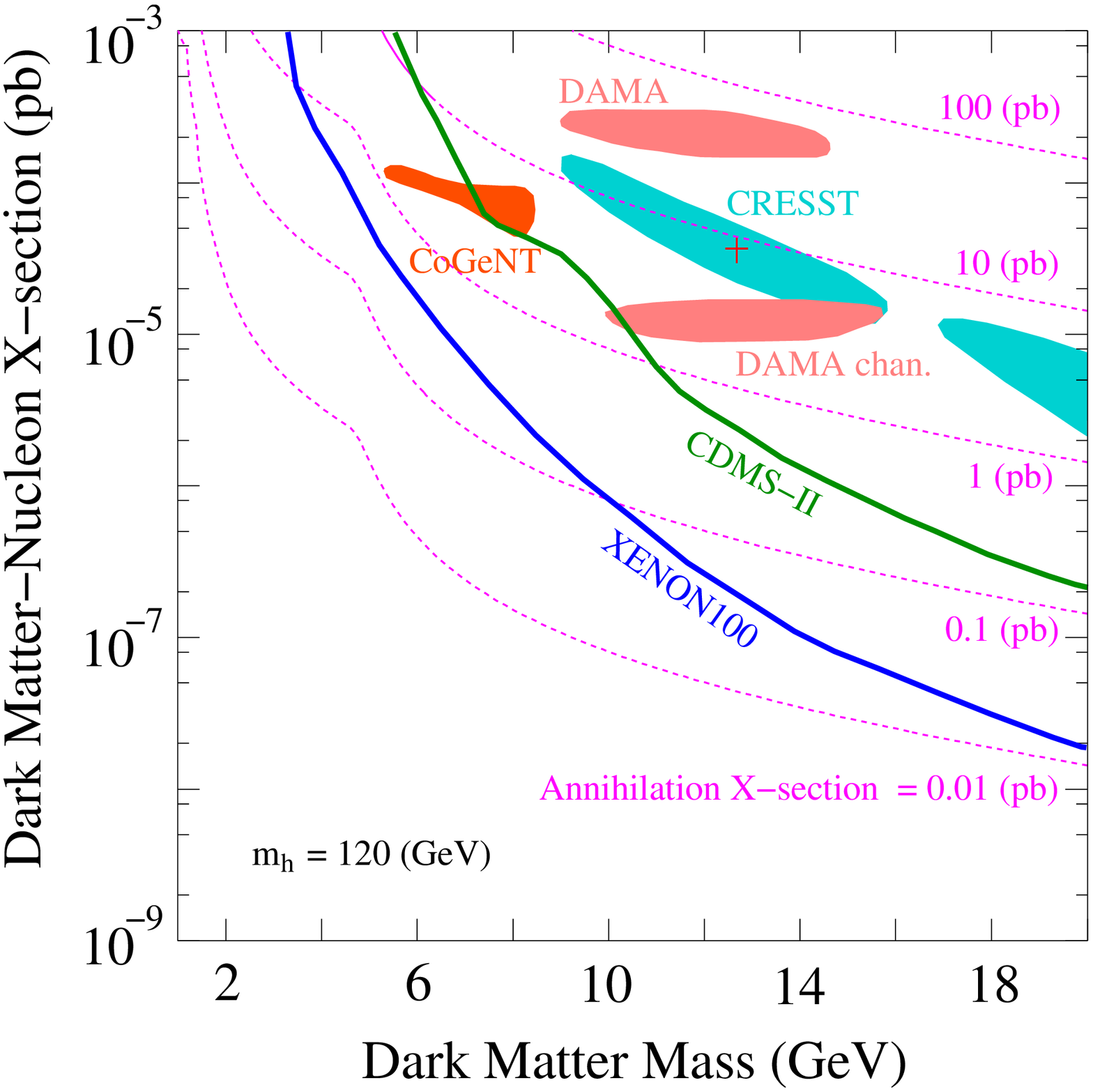}
\caption{\small
(Left panel) The annihilation cross section between the bosonic dark and anti-dark matter particles as a function of the dark matter for $\kappa_H = 1$. We have taken the threshold of the bottom mode to be $m_B = 5.27$ GeV. (Right panel) The scattering cross section between the bosonic dark matter and a nucleon as a function of the dark matter mass for a given annihilation cross section. We also show results reported by DAMA~\cite{Savage:2008er}, CoGeNT~\cite{CoGeNT}, and CRESST~\cite{Angloher:2011uu} collaborations. The upper limits on the cross section reported by XENON100~\cite{Aprile:2011hi} and CDMS-II~\cite{Ahmed:2009zw} collaborations are also shown.}
\label{fig: bosonic}
\end{center}
\end{figure}

In order to eliminate the symmetric component of dark matter relics in our universe, the annihilation cross section between dark and anti-dark matter particles has to be large enough as mentioned before. For the bosonic dark matter, we have the following interaction in general, which is renormalizable and symmetric,
\begin{eqnarray}
{\cal L}_{\rm ann} = \kappa_H |\phi|^2|H|^2.
\label{eq: ann}
\end{eqnarray}
With this interaction as well as SM interactions, the annihilation cross section (times the relative velocity $v$) for the bosonic dark matter is obtained to be,
\begin{eqnarray}
\sigma v
=
\sum_f \frac{c_f \kappa_H^2 m_f^2}{4 \pi (m_h^2 - s)^2}
\left(1 - \frac{4m_f^2}{s}\right)^{3/2}
+
\frac{\alpha_s^2(m) \kappa_H^2 s}{16 \pi^3 (m_h^2 - s)^2}
\left|\sum_q F(4m_q^2/s) \right|^2,
\label{eq: annCS}
\end{eqnarray}
where $s$, $c_f$, and $\alpha_s(m)$ are the center of mass energy, the color factor of each fermion, and the strong coupling constant evaluated at the scale $m$, respectively. The notation $f$ denotes the fermion ($f = u, d, s, c, b, e, \mu, \tau$), while $f_c$ is the quark ($q = u, d, s, c, b, t$). The first contribution in the r.h.s. of the equation comes from the annihilation into a pair of fermions with the $s$-channel higgs exchange, while the second one is from the annihilation into two gluons also with the higgs exchange through the one-loop triangle diagram. Here, the loop function $F$ is given by $F(x) = x + x(1 - x)(\sin^{-1}\sqrt{1/x})^2$. The annihilation cross section as a function of the dark matter mass $m$ is shown in Fig.\ref{fig: bosonic} (the left panel) with $s$ being $4m^2$. It can be seen that the cross section can be 1--10 pb when $m =$ 5--10 GeV, which is large enough to eliminate the symmetric component of dark matter relics~\cite{Graesser:2011wi}.

The renormalizable interaction in eq.(\ref{eq: ann}) also provides a rather large cross section between the bosonic dark matter and a nuclei, which gives a significant impact on direct detection experiments of dark matter. In fact, through the $t$-channel exchange of the higgs bosons, the bosonic dark matter is expected to interact with a nuclei inside the detector, whose cross section per nucleon $N$ is estimated to be
\begin{eqnarray}
\sigma_N \simeq
\frac{\kappa_H^2}{4 \pi m_h^4} \frac{m_N^2}{(m_{DM} + m_N )^2} f_N^2,
\end{eqnarray}
where $f_N$ is the coupling between the dark matter and a nucleon $N$, which is estimated to be $f_N \simeq 0.266 \, m_N$ with $m_N$ being the nucleon mass~\cite{Kanemura:2010sh}. In Fig.\ref{fig: bosonic} (the right panel), we show the scattering cross section between the dark matter and a nucleon as a function of the dark matter mass for a given annihilation cross section. The figure shows that the bosonic dark matter with the annihilation cross section of 1--10 pb and the mass around 5--7 GeV is remarkably consistent with the CoGeNT anomaly~\cite{CoGeNT}. In the figure, we also show upper limits on the scattering cross section reported by Xenon100~\cite{Aprile:2011hi} and CDMS-II~\cite{Ahmed:2009zw} experiments, showing a tension with the CoGeNT anomaly. Since the consistency among these experiments with involving astrophysical and detector uncertainties is still under debate, we do not go more details on this issue, and leave for discussions of other papers~\cite{Uncertainties}.

Another interesting prediction derived from the renormalizable interaction in eq.(\ref{eq: ann}) is the invisible decay of the higgs boson into a pair of dark matter particles, $h \to \phi + \phi^*$. In fact, its decay width is dominated by the invisible mode for $\kappa_H = {\cal O}(1)$. The higgs boson will therefore slip through the current search at the LHC which is based on the branching ratio into SM particles~\cite{LHC}. Discovery potential of the invisibly decaying higgs at the LHC via the $W$-boson fusion has been discussed in refs.~\cite{InvH} for the 14 TeV run. Notice that scalar WIMP dark matter models for $m =$ 3--20 GeV have been severely constrained by observations of cosmic-ray anti-protons~\cite{Kappl:2011jw} such as BESS-Polar~\cite{BESS} and PAMELA satellite~\cite{PAMELA} experiments, if the annihilation cross section is provided by the renormalizable interaction. This is because the WIMP dark matter mainly annihilates into a pair of bottom quarks at the present universe, which leads to the too high anti-proton flux for $\sigma v \simeq 1$ pb. The asymmetric dark matter, on the other hand, does not self-annihilate at the present universe due to the asymmetry, and hence free from this constraint.

Before closing this subsection, let us comment on a serious drawback of the model in eq.(\ref{eq: bosonic}). Since the B$-$L charge of the dark matter is identical to the one of right-handed neutrinos, the large $\Delta({\rm B-L}) =2$ mass term of the dark matter is expected to be generated once right-handed neutrinos acquire their mass terms,
\begin{eqnarray}
\label{eq: BL2}
{\cal L}
= m_{\rm LV}^2 \phi^2 + h.c.
\qquad
{\rm with}~~~
m_{\rm LV}^2 \simeq M_{*} M_{N},
\end{eqnarray}
where $M_N$ and $M_{*}$ denote the typical mass of right-handed neutrinos and the cutoff scale of the theory such as the Planck scale, respectively. As discussed in refs.~\cite{Regeneration}, however, the $\Delta({\rm B-L}) =2$ mass term causes the oscillation between dark and anti-dark matter particles over the age of the universe, which erases the asymmetry of dark matter relics. Once the asymmetry is erased, the dark matter can annihilate into SM particles with a larger cross section than those of WIMP dark matters, which contradicts with constraints obtained from observations of nearby dwarf spheroidal galaxies with the Fermi Gamma-Ray Space Telescope~\cite{collaboration:2011wa}. As a result, $m$ is required to be as small as $10^{-41}$ GeV. We therefore need a severe fine-tuning to satisfy this condition in the model defined in eq.(\ref{eq: bosonic}). Furthermore, even if we forbid the $\Delta({\rm B-L})=2$ mass term in eq.(\ref{eq: BL2}) by hand, it is generated again by three-loop diagrams, because the operator ${\cal O}_{SM}$ can annihilate into the mass terms of right-handed neutrinos. This radiatively generated $\Delta({\rm B-L})=2$ mass term is still much larger than the above constraint, which also require a severe fine-tuning.

In order to avoid this problem, we may consider an additional symmetry such as the so-called Peccei-Quinn symmetry by extending the higgs sector to the two higgs doublet model. In this case, the bosonic dark matter as well as the higgs bosons carries new charges consistent with the interaction in eq.(\ref{eq: bosonic}), while the $\Delta({\rm B-L})=2$ mass term is forbidden.\footnote{A rather large discrete symmetry after the breaking of the Peccei-Quinn symmetry is actually needed in order to forbid the dangerous $\Delta({\rm B-L})=2$ mass term.} Another way to solve this problem is the use of other operators for ${\cal O}_{SM}$ instead of $(LH)^2$. In particular, when ${\cal O}_{SM}$ carries a non-trivial B number such as ${\cal O}_{SM} = (u_Rd_Rd_R)^2$, the dark matter carries the B charge, and the $\Delta({\rm B-L})=2$ mass term is not expected even after the B$-$L breaking. Here, we implicitly assume that the UV completed theory inducing the interaction $\phi^2{\cal O}_{SM}$ has the B symmetry as an accidental symmetry as it happens in the SM.

Finally, we comment on another constraint on the bosonic dark matter. As discussed in refs.~\cite{BH constraints}, once bosonic asymmetric dark matters are accumulated in neutron stars, they form black holes because of the absence of their self-annihilations. This fact leads to a sever upper bound on the scattering cross section between the dark matter and a nucleon. We point out, however, that it may be possible that the dark matter can annihilate in an asymmetric manner through the ADM interaction in Eq.\,(\ref{eq: bosonic}) when $\Lambda$ is low enough~\cite{Asymmetric annihilation}. Such a bosonic asymmetric dark matter model can evade the constraint from the black hole formation inside neutron stars. Furthermore, it might be even possible to search the asymmetric dark matter at neutrino detectors such as Super-Kamikande~\cite{SK} by looking for the line spectrum of neutrinos emitted from asymmetric annihilations. More ambitiously, if we can distinguish the neutrino line from the anti-neutrino line, it will provide us important hints on the asymmetry of the dark matter and the neutrino sectors.

\subsection{Fermionic dark matter model}

Next, let us consider the model for the fermionic dark matter with the lowest dimensional connection. Interestingly, the operator ${\cal O}_{\rm SM}$ can be not only bosonic but also fermionic such as ${\cal O}_{\rm SM} = LH$ in this case. The lowest dimensional operator which connects between the dark matter particle and SM particles turns out to be
\begin{eqnarray}
\label{eq: fermionic}
{\cal L}_{\rm int}
=
\frac{1}{\Lambda^3} \chi^3 \cdot (LH) + h.c.,
\end{eqnarray}
where the B$-$L charge of the dark matter is assumed to be one third. According to the generic argument in previous section, the mass of the dark matter is thus predicted to be $m \simeq$ 15--21 GeV in this case. One of the advantages to consider this model compared to the case of the bosonic dark matter model in previous subsection is that the $\Delta({\rm B-L}) =2$ mass term is not allowed even after the spontaneous symmetry breaking of $U(1)_{\rm B-L}$. This is because, even after the breaking, we have a residual $Z_6$ symmetry, so that the dangerous $\Delta({\rm B-L}) =2$ mass term is prohibited. Furthermore, we do not have serious problems on the constraint coming from neutron stars even if the new physics scales $\Lambda$ is large, because the fermionic dark matter does not accumulated thanks to the Pauli Exclusion Principle.

Finally, let us briefly discuss a UV completed model (composed only of renormalizable interactions) which derives the connection in eq.(\ref{eq: fermionic}). The simplest model is, for example, possible to be constructed by introducing a SU(2)$_L$ doublet scalar ($s_d$), and a pair of SU(2)$_L$ doublet fermions $(\psi_d, \bar{\psi}_d)$ with the Lagrangian, 
\begin{eqnarray}
{\cal L}_{\rm UV}
&=&
{\cal L}_{\rm K}
+
\left(
\lambda_L \bar{L} s_d \chi^c  
+\lambda_H \bar{\psi}_d H^c \chi^c
+\lambda_\psi \bar{\psi}_d s_d \chi + h.c.
\right),
\nonumber \\
{\cal L}_{\rm K}
&=&
{\cal L}_{\rm kin}
-m \bar{\chi} \chi
-m_s^2 |s_d|^2
-m_f \bar{\psi}_d \psi_d,
\label{eq: UV completion}
\end{eqnarray}
where the superscript '$c$' denotes charge conjugation, and ${\cal L}_{\rm kin}$ is composed of kinetic terms of $\chi$($\bar{\chi}$), $s_d$, and $\psi$($\bar{\psi}$) fields. Here, we have assumed that all coupling constants ($\lambda_L$, $\lambda_L$, and $\lambda_\phi$) take real values for simplicity. Gauge quantum numbers of these fields are given in Table.\ref{tab: quantum numbers 2}. After integrating the doublet scalar ($s_d$) and fermions $(\psi_d, \bar{\psi}_d)$ out from the Lagrangian, we obtain following effective interactions,
\begin{eqnarray}
{\cal L}_{\rm eff}
=
\frac{\lambda_H^2}{m_f} |H|^2 (\bar{\chi}\chi + h.c.)
+\frac{\lambda_L^2}{m_s^2}|\bar{L} \chi|^2
+\frac{\lambda_L \lambda_\psi \lambda_H}{m_f m_s^2} 
\left[
(\bar{\chi^c}\chi)(\bar{\chi^c} L) H + h.c.
\right],
\label{eq: effective interactions}
\end{eqnarray}
up to dimension-seven operators. Though we have other higher-dimensional operators, they are irrelevant for our discussions and we thus omit to wright them.

\begin{table}[t]
\centering
\begin{tabular}{c|ccccccc}
& $\chi$ & $\bar\chi$&$s_d$ & $\psi_d$ & $\bar\psi_d$  \\
\hline
SU(3)$_c$ & {\bf{1}} & {\bf 1} & {\bf 1} & {\bf 1} & {\bf 1}  \\
SU(2)$_L$ & {\bf 1} & {\bf 1} & {\bf 2} & {\bf 2} & {\bf 2} \\
U(1)$_Y$ & $0$ & 0&$-1/2$ & $-1/2$ & $1/2$ \\
U(1)$_{\rm B-L}$ & $1/3$ & $-1/3$&$-2/3$ & $-1/3$ & $1/3$   \\
\hline
\end{tabular}
\caption{\small Quantum numbers of fields in the UV completion.}
\label{tab: quantum numbers 2}
\end{table}

The renormalizable model in eq.(\ref{eq: UV completion}) therefore leads to the B$-$L connecting interaction in eq.(\ref{eq: fermionic}) as the third term by identifying the coefficient to be
\begin{eqnarray}
\frac{1}{\Lambda^3} = 
\frac{\lambda_L \lambda_\psi \lambda_H}{m_f m_s^2}.
\end{eqnarray}

Finally, we comment on the annihilation cross section between dark and anti-dark matter particles in this model. Since the annihilation through the first operator in eq.~(\ref{eq: effective interactions}) turns out to be velocity suppressed, it is difficult to eliminate the symmetric component of dark matter relics from our universe. Therefore, as discussed in ref.~\cite{Kaplan:2009ag}, we should extend the higgs sector of the SM to obtain the required annihilation cross section, $\sigma v =$ 1-10 pb. For example, by adding a singlet scalar field into the higgs sector, we can easily obtain the annihilation cross section of the order of 10 pb.

\section{Conclusions and Discussion}
\label{sec: conclusion}

We have discussed the dark matter whose stability is guaranteed by a residual discrete symmetry of U(1)$_{\rm B-L}$. We have especially considered a class of models where there is new physics connecting B$-$L charges of the dark matter to those of SM particles at the energy scale between electroweak and U(1)$_{\rm B-L}$ breaking. In this class of models, the ratio of the dark matter to the baryon relic densities can be solely determined by the mass and B$-$L charge of the dark matter, when the baryon asymmetry of the universe is generated through the leptogenesis.\footnote{Our analysis is also possible to apply to the cases that the B$-$L asymmetry is mainly generated not in the SM sector but in the dark matter sector by some other mechanism rather than the usual leptogenesis, although such new mechanism requires more additional structures~\cite{Darkogenesis}.} As a result, we have proved that the mass of the dark matter is definitely predicted to be (5--7)/$Q_{\rm DM}$ without depending on details of new physics models and also its energy scale.

The mass of dark matter is measurable at direct detection experiments of dark matter. We can therefore probe the dark matter scenario discussed in this article through its mass measurement, which, in turn, provides us indirect but important hints on the leptogenesis as the origin of the baryon asymmetry of the universe.

\section*{Acknowledgments}

This work is supported by Grant-in-Aid for Scientific research from the Ministry of Education, Science, Sports, and Culture (MEXT), Japan, No.\ 22244021 (S.M. and T.T.Y.) and No.\ 23740169 (S.M.), and also by World Premier International Research Center Initiative (WPI Initiative), MEXT, Japan.


\begin{thebibliography}{99}

\bibitem{Komatsu:2010fb}
  E.~Komatsu {\it et al.}  [WMAP Collaboration],
  Astrophys.\ J.\ Suppl.\  {\bf 192}, 18 (2011).

\bibitem{Evrard:2001hu}
  See, for example,
  A.~E.~Evrard {\it et al.}  [The VIRGO Collaboration],
  Astrophys.\ J.\  {\bf 573}, 7 (2002).

\bibitem{Nakayama:2011dj}
  K.~Nakayama, F.~Takahashi, T.~T.~Yanagida,
  Phys.\ Lett.\  {\bf B699}, 360-363 (2011).

\bibitem{seesaw}
  T.~Yanagida, in Proceedings of the Workshop on Unified Theories
  and Baryon Number in the Universe, eds.\ O.~Sawada and A.~Sugamoto
  (KEK report 79-18, 1979);
  M.~Gell-Mann, P.~Ramond and R.~Slansky, 
  in Sanibel Symposium, Palm Coast, Fla., Feb 1979 (hep-ph/9809459);
  see also
  P.~Minkowski,
  Phys.\ Lett.\  B {\bf 67}, 421 (1977).

\bibitem{leptogenesis}
  M.~Fukugita and T.~Yanagida,
  Phys.\ Lett.\  B {\bf 174}, 45 (1986);
  W.~Buchmuller, R.~D.~Peccei and T.~Yanagida,
  Ann.\ Rev.\ Nucl.\ Part.\ Sci.\  {\bf 55}, 311 (2005);
  S.~Davidson, E.~Nardi and Y.~Nir,
  ``Leptogenesis,''
  Phys.\ Rept.\  {\bf 466}, 105 (2008).

\bibitem{Darkogenesis}
  J.~Shelton and K.~M.~Zurek,
  Phys.\ Rev.\  D {\bf 82}, 123512 (2010);
  N.~Haba and S.~Matsumoto,
  Prog.\ Theor.\ Phys.\  {\bf 125}, 1311 (2011).

\bibitem{CoGeNT}
  C.~E.~Aalseth {\it et al.},
  Phys.\ Rev.\ Lett.\  {\bf 107}, 141301 (2011);
  C.~E.~Aalseth {\it et al.}  [CoGeNT collaboration],
  Phys.\ Rev.\ Lett.\  {\bf 106}, 131301 (2011).

\bibitem{Kaplan:2009ag}
  D.~E.~Kaplan, M.~A.~Luty and K.~M.~Zurek,
  Phys.\ Rev.\  D {\bf 79}, 115016 (2009).

\bibitem{AristizabalSierra:2010mv}
  D.~Aristizabal Sierra, J.~F.~Kamenik and M.~Nemevsek,
  JHEP {\bf 1010}, 036 (2010).

\bibitem{Harvey:1990qw}
  See, for example,
  J.~A.~Harvey and M.~S.~Turner,
  Phys.\ Rev.\  D {\bf 42}, 3344 (1990).

\bibitem{Nakamura:2010zzi}
  K.~Nakamura {\it et al.}  [Particle Data Group],
  J.\ Phys.\ G {\bf 37}, 075021 (2010).

\bibitem{Buckley:2010ui}
  M.~R.~Buckley and L.~Randall,
  JHEP {\bf 1109}, 009 (2011).

\bibitem{Graesser:2011wi}
  M.~L.~Graesser, I.~M.~Shoemaker, L.~Vecchi,
  [arXiv:1103.2771 [hep-ph]].

\bibitem{Kanemura:2010sh}
  See, for example,
  S.~Kanemura, S.~Matsumoto, T.~Nabeshima, N.~Okada,
  Phys.\ Rev.\  {\bf D82}, 055026 (2010),
  and references therein.

\bibitem{Savage:2008er}
  C.~Savage, G.~Gelmini, P.~Gondolo and K.~Freese,
  JCAP {\bf 0904}, 010 (2009).

\bibitem{Angloher:2011uu}
  G.~Angloher {\it et al.},
  arXiv:1109.0702 [astro-ph.CO].

\bibitem{Aprile:2011hi}
  E.~Aprile {\it et al.}  [XENON100 Collaboration],
  arXiv:1104.2549 [astro-ph.CO].

\bibitem{Ahmed:2009zw}
  Z.~Ahmed {\it et al.}  [The CDMS-II Collaboration],
  Science {\bf 327}, 1619 (2010).

\bibitem{Uncertainties}
  For a latest analysis, see, for example,
  C.~Arina,
  arXiv:1110.0313 [astro-ph.CO];
  C.~Arina, J.~Hamann and Y.~Y.~Y.~Wong,
  JCAP {\bf 1109}, 022 (2011).

\bibitem{LHC}
  \verb$http://lhc.web.cern.ch/lhc/$.

\bibitem{InvH}
  O.~J.~P.~Eboli and D.~Zeppenfeld,
  Phys.\ Lett.\  B {\bf 495}, 147 (2000);
  M.~Warsinsky  [ATLAS Collaboration],
  J.\ Phys.\ Conf.\ Ser.\  {\bf 110}, 072046 (2008);
  Di~Girolamo~B and Neukermans~L, Atlas Note ATL-PHYS-2003-006 (2003).

\bibitem{Kappl:2011jw}
  R.~Kappl and M.~W.~Winkler,
  arXiv:1110.4376 [hep-ph].

\bibitem{BESS}
  K.~Abe {\it et al.},
  arXiv:1107.6000 [astro-ph.HE].

\bibitem{PAMELA}
  O.~Adriani {\it et al.}  [PAMELA Collaboration],
  Phys.\ Rev.\ Lett.\  {\bf 105}, 121101 (2010).

\bibitem{Regeneration}
  M.~R.~Buckley and S.~Profumo,
  arXiv:1109.2164 [hep-ph];
  M.~Cirelli, P.~Panci, G.~Servant and G.~Zaharijas,
  arXiv:1110.3809 [hep-ph].

\bibitem{collaboration:2011wa}
  T.~L.~collaboration,
  arXiv:1108.3546 [astro-ph.HE].

\bibitem{BH constraints}
  S.~D.~McDermott, H.~B.~Yu and K.~M.~Zurek,
  arXiv:1103.5472 [hep-ph];
  C.~Kouvaris and P.~Tinyakov,
  Phys.\ Rev.\ Lett.\  {\bf 107}, 091301 (2011).

\bibitem{Asymmetric annihilation}
A  M.~Ibe, S.~Matsumoto and T.~T.~Yanagida, in preparation.

\bibitem{SK}
  \verb$http://www-sk.icrr.u-tokyo.ac.jp/sk/index-e.html$.

\end{thebibliography}
\end{document}